\documentclass[conference, 9pt]{IEEEtran}
\usepackage{amsmath}
\usepackage{amsfonts}
\usepackage{amssymb}

\newtheorem{definition}{Definition}
\newtheorem{lemma}{Lemma}
\newtheorem{theorem}{Theorem}
\newtheorem{proposition}{Proposition}

\usepackage{tikz}
\usetikzlibrary{decorations.markings}
\tikzset{->-/.style={decoration={markings, mark=at position .5 with {\arrow{>}}}, postaction={decorate}}}
\usetikzlibrary{shapes.geometric}

\usepackage{xcolor}

\usepackage{hyperref}
\newcommand\G{\mathcal G} 
\renewcommand\H{\mathcal H} 
\newcommand\D{\mathcal D} 
\renewcommand\P{\mathbf P} 
\newcommand\C{\mathbf C} 
\newcommand\B{\mathbf B} 
\newcommand\RR{\mathbb R} 
\newcommand\id{\mathrm{id}}
\newcommand\op{\mathrm{op}}
\DeclareMathOperator*\argmax{\arg\max}
\newcommand\Set{\mathbf{Set}}
\renewcommand\Game{\mathbf{Game}}
\DeclareMathOperator\Rel{Rel} 
\title{Compositional Game Theory}
\author{Neil Ghani \and Jules Hedges \and Viktor Winschel \and Philipp Zahn}
\begin{document}
\maketitle

\begin{abstract}
We introduce open games as a compositional foundation of economic game theory. A compositional approach potentially allows methods of game theory and theoretical computer science to be applied to large-scale economic models for which standard economic tools are not practical. An open game represents a game played relative to an arbitrary environment and to this end we introduce the concept of coutility, which is the utility generated by an open game and returned to its environment. 
Open games are the morphisms of a symmetric monoidal category and 
can therefore be composed by categorical composition into sequential move games and
by monoidal products into simultaneous move games.
Open games can be represented by string diagrams which provide an intuitive but formal visualisation of the information flows. 
We show that a variety of games can be faithfully represented as open games in the sense of having the same Nash equilibria and off-equilibrium best responses.
\end{abstract}

\section{Introduction}\label{sec:introduction}
The concept of \emph{compositionality} is well-known and almost commonplace in computer science, 
where it is what ultimately allows programmers to scale software to large systems.
However, in many other fields compositionality is essentially unknown and hence its benefits are not available. In this paper we introduce compositionality into a field where one might not believe it to be possible: the study of strategic games and Nash equilibria. They are of interest in economics and computer science where optimal decisions are taken by interacting agents with conflicting goals.\footnote{Games in the sense of \emph{game semantics} are compositional, but they avoid several difficult defining features of game theory by restricting to the 2-player zero-sum setting.}

In contrast to classical game theory, where games are studied monolithically as one global object, compositional game theory works bottom-up by building large and complex games from smaller components. Such an approach is inherently difficult since the interaction between games has to be considered. Moreover, in the compositional approach, the equilibria of larger games should be defined from the equilibria of the component games - but {\em a priori}, there is no reason why this should be possible.  

For example, in the prisoner's dilemma game, each player's best option is to
defect, although, if they acted as a single agent, they would cooperate.
Moreover, if the one-shot prisoner's dilemma game is repeated, then cooperative equilibria become achievable. More generally, the equilibria of a composite game are not necessarily made up from those of the component games, and locally optimal moves are not guaranteed to be globally optimal. In essence, game theory contains  \emph{emergent effects} whereby a composite system exhibits behaviours that are not (simple) functions of the behaviours of the components. Accordingly, emergent effects make compositionality very hard to achieve and the existence of a compositional model of game theory is somewhat surprising. In order to arrive at this goal we had to radically reformulate classical game theory from first principles and rebuild it on \emph{open games}. 

Open games represent the relationship between different interactions in two dimensions: in sequence, if an interaction follows another interaction and in parallel, if interactions take place simultaneously. As such, we follow a path taken in the  field of \emph{open systems} \cite{willems_behavioural_approach_open_interconnected_systems}, and in particular \emph{categorical open systems} \cite{fong_algebra_open_interconnected_systems} where compositional approaches to general systems are studied.  Here, systems are modelled as morphisms $f : X \to Y$ in a symmetric monoidal category, where the objects $X$ and $Y$ describe the \emph{boundaries} of the open system, where it interacts with its environment. This means that systems $f : X \to Y$ and $f' : X' \to Y'$ can be composed in parallel using the monoidal product to yield $f \otimes f' : X \otimes X' \to Y \otimes Y'$, and two systems $f : X \to Y$ and $g : Y \to Z$ sharing a common boundary can be composed sequentially by glueing along this boundary to yield $g \circ f : X \to Z$. Ordinary, closed systems are recovered as scalars \cite{abramsky05}, i.e. endomorphisms $f : I \to I$ of the monoidal unit, which represents a trivial boundary. Open games are accordingly the morphisms of a symmetric monoidal category.

A compositional model of game theory does not only have to model a game but also
the interactions of the game with all other games and environments. This can be
seen as a form of \emph{continuation passing style}. This would still be hardly
tractable if the environment of an open game included arbitrary other open games. The crucial technical feature underlying our approach is to describe the behaviour of an open game relative to a simplified notion of an environment which we call a \emph{context}, in which the future is abstracted into a single utility function.  In this way, we reduce an arbitrarily complex game to a set of individual decisions. The circularity of a Nash equilibrium, where all players play mutually best replies, is finally handled by the composition operators.

The theory of open games is based on two main predecessors. Firstly, in \cite{pavlovic09} games are defined as processes and in \cite{blumensath13} the dynamics but not the equilibria are treated compositionally. The second predecessor is the theory of higher order games and selection functions, for example in \cite{escardo11} and \cite{hedges_etal_selection_equilibria_higher_order_games}, which give a theory of equilibria relative to an environment but are not strongly compositional. 
Selection functions can be used to model \emph{goals} of agents compositionally \cite{hedges_etal_higher_order_decision_theory}.
Combining features of these approaches into a single mathematical object required the innovations mentioned above and led us to discover the idea of an open game. 
After we developed open games, connections to lenses and the geometry of interaction were noticed respectively by Jeremy Gibbons and Tom Hirschowitz.

We omit proofs in this paper, which can be found in \cite{hedges_towards_compositional_game_theory} and \cite{hedges_morphisms_open_games}. We also work over the category of sets to keep notation and overheads to a minimum -- for a full categorical account, see once more \cite{hedges_towards_compositional_game_theory}.
The rest of this paper is structured as follows: The next section introduces selection functions as a key ingredient to open games. Section \ref{sec:open-games} introduces the definition of an open game and discusses its elements, followed by some examples in Section  \ref{sec:examples}. The monoidal category of open games is introduced in Section \ref{sec:category} and the string diagrams attached to this category in Section \ref{sec:string_diagram}. We then turn to examples built compositionally: in Section \ref{sec:simultaneous} we discuss simultaneous move games and in Section \ref{sec:sequential} sequential move games. Section \ref{sec:conclusion} concludes the paper with an outlook on further work. 

\section{Selection functions and higher order games}\label{sec:games}
For reasons of space, we assume the reader knows some basic
game theory, such as the definitions of normal-form and extensive-form games and
Nash equilibrium. These basic concepts can be found for example in \cite{brown08} or many online lecture notes. Neverthless, in this section we 
introduce enough game theory via selection functions~\cite{escardo11,hedges_etal_selection_equilibria_higher_order_games}
 to make the paper self contained.

\begin{definition}
	An $n$-player \emph{higher order simultaneous move game} is defined by the following data:
	\begin{itemize}
		\item Sets $X_1, \ldots, X_n$ of \emph{choices} for each player
		\item A set $R$ of \emph{outcomes}
		\item An \emph{outcome function} $q : \prod_{i = 1}^n X_i \to R$
		\item For each player $1 \leq i \leq n$, a \emph{multi-valued selection function} $\delta_i : (X_i \to R) \to \mathcal P (X_i)$
	\end{itemize}
\end{definition}

In this game, each player simultaneously makes a choice of move $x_i : X_i$.
We define a (pure) \emph{strategy} for player $i$ to be just a choice in $X_i$, and a (pure) \emph{strategy profile} to be a tuple of strategies in $\Sigma := \prod_{i = 1}^n X_i$.
When all choices are made, the rules of the game determine an outcome $q (x_1, \ldots, x_n) : R$. The selection function $\delta_i : (X_i \to R) \to \mathcal P (X_i)$ defines the set of moves $\delta_i (k)$ that are considered optimal in a \emph{context} $k : X_i \to R$, which describes the individual decision faced by player $i$. A context for player $i$ is obtained from the outcome function by fixing strategies for all other players and is thus the utility function associating, to each potential unilateral deviation $x_i$ of player $i$, the utility arising from that deviation.

Given a higher order simultaneous move game, we define its \emph{best response relation} $\B \subseteq \Sigma \times \Sigma$ by $(\sigma, \sigma') \in \B$ iff for all players $1 \leq i \leq n$, we have
\[\sigma'_i \in \delta_i (\lambda (x_i : X_i) . q (\sigma [i \mapsto x_i])) \]
where $\sigma [i \mapsto x_i]$ is the strategy profile $\sigma$ apart from the $i$th player who chooses $x_i$ or, more formally, 
\[ (\sigma [i \mapsto x_i])_j = \begin{cases}
	x_i &\text{ if } i = j \\
	\sigma_j &\text{ otherwise}
\end{cases} \]
A \emph{selection equilibrium} is a pure strategy profile $\sigma$ with $(\sigma, \sigma) \in \B$.

The classical definition of an $n$-player simultaneous move game in normal form results from this definition as the special case in which the set of outcomes is $\RR^n$ and the $i$th player's selection function is
\begin{align*}
	\delta_i (k) &= \argmax (\pi_i \circ k) \\
	&= \{ x : X_i \mid (k (x))_i \geq (k (x'))_i \text{ for all } x' : X_i \}
\end{align*}
In this case the selection equilibria agree with the usual definition of pure strategy Nash equilibrium, and moreover the best response relation is the same.
There are many well-known examples of 2-player simultaneous move games with 2 moves each, such as the prisoner's dilemma, matching pennies, battle of the sexes, chicken, etc., defined by different outcome functions $q : \{ C, D \}^2 \to \RR^2$.
The prisoner's dilemma, for example, is given by the outcome function
\begin{align*}
	&q (C, C) = (2, 2) &&q (C, D) = (0, 3) \\
	&q (D, C) = (3, 0) &&q (D, D) = (1, 1)
\end{align*}
and yields the constant best response relation with $(\sigma, (D, D)) \in \B$ for all strategy profiles $\sigma$.
This happens because $D$ is a \emph{dominant strategy} for both players in the prisoner's dilemma.
An extended example can be found in
\cite{hedges_etal_selection_equilibria_higher_order_games} of a higher order
simultaneous move game whose selection functions are not of this form, which we will discuss in section \ref{sec:simultaneous}.

\begin{definition}
	An $n$-player \emph{higher order sequential game} is defined by the same data as a simultaneous move game: sets $X_1, \ldots, X_n$ of choices, a set $R$ of outcomes, an outcome function $q : \prod_{i = 1}^n X_i \to R$, and selection functions $\delta_i : (X_i \to R) \to \mathcal P (X_i)$.
	A strategy for player $i$ is a function
	\[ \sigma_i : \prod_{j = 1}^{i - 1} X_j \to X_i \]
	that chooses a move contingent on the previous moves by other players, and the set of strategy profiles is the set
	\[ \Sigma = \prod_{i = 1}^n \left( \prod_{j = 1}^{i - 1} X_j \to X_i \right) \]
	of tuples consisting of a strategy for each player.
	There is an obvious play function
	\[ \P : \Sigma \to \prod_{i = 1}^n X_i \]
	producing the sequence of moves resulting from a strategy profile, defined by course-of-values recursion, which in the base case uses
	\[ \prod_{j = 1}^0 X_j \to X_1 \cong X_1 \]
	Given a strategy profile $\sigma$, we call $\P (\sigma)$ the \emph{strategic play} of $\sigma$.
\end{definition}

The best response relation $\B \subseteq \Sigma \times \Sigma$ of a higher order sequential game is defined by $(\sigma, \sigma') \in \B$ iff
\[ (\mathbf P (\sigma [i \mapsto \sigma'_i]))_i \in \delta_i (\lambda (x_i : X_i) . q (\mathbf U_i (x_i, \sigma))) \]
where the \emph{unilateral deviation operator} $\mathbf U_i$ is the evident function
\[ \mathbf U_i : X_i \times \Sigma \to \prod_{j = 1}^n X_j \]
defined by course-of-values recursion, with $(\mathbf U_i (x_i, \sigma))_i = x_i$.
Strategy profiles $\sigma$ with $(\sigma, \sigma) \in \B$ are called \emph{selection equilibria} of the sequential game.

In the case where the selection functions are $\argmax$ as defined above, this agrees with the standard definitions of best responses and Nash equilibria.
Note that this is a strictly weaker definition than that of \emph{optimal strategies} from \cite{escardo11}, which specialises to \emph{subgame perfect equilibrium} (a strengthening of Nash equilibrium) when the selection functions are $\argmax$ \cite{escardo12}.

\section{Open games}\label{sec:open-games}

The primary objects of study in compositional game theory are called \emph{open games}.
We start by giving the definition, and in the remainder of this section we discuss its interpretation. 

\begin{definition}
	Let $X, S, Y, R$ be sets.
	An \emph{open game} $\G : (X, S) \to (Y, R)$ is defined to be a 4-tuple $\G = (\Sigma_\G, \P_\G, \C_\G, \B_\G)$, where
	\begin{itemize}
		\item $\Sigma_\G$ is a set, called the set of \emph{strategy profiles} of $\G$
		\item $\P_\G : \Sigma_\G \times X \to Y$ is called the \emph{play function} of $\G$
		\item $\C_\G : \Sigma_\G \times X \times R \to S$ is called the \emph{coplay function} of $\G$
		\item $\B_\G : X \times (Y \to R) \to \operatorname{Rel} (\Sigma_\G)$ is called the \emph{best response function} of $\G$
	\end{itemize}
\end{definition}
$\operatorname{Rel} (\Sigma_\G)$ is the meet-semilattice of all endo-relations $R \subseteq \Sigma_\G \times \Sigma_\G$.
In general, we impose no conditions whatsoever on these components. In practice, however, we are most interested in those open games which are generated by certain constructions, some of which are defined in this paper. There are class-many open games of a fixed type (which causes the category of open games to be locally large), but only set-many after restricting to those generated by a set of constructions.
We will represent a general open game $\G : (X, S) \to (Y, R)$ in diagrammatic form (where time flows from left to right) as
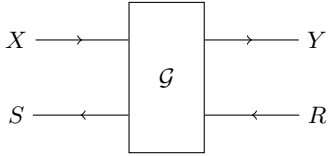
\begin{center} \begin{tikzpicture}
	\node (X) at (-2, .5) {$X$}; \node (Y) at (2, .5) {$Y$}; \node (R) at (2, -.5) {$R$}; \node (S) at (-2, -.5) {$S$};
	\node [rectangle, minimum height=2cm, minimum width=1cm, draw] (G) at (0, 0) {$\G$};
	\draw [->-] (X) to (G.west |- X); \draw [->-] (G.east |- Y) to (Y); \draw [->-] (R) to (G.east |- R); \draw [->-] (G.west |- S) to (S);
\end{tikzpicture} \end{center}
In Section \ref{sec:category} we will build on this and discuss the graphical syntax of open games.
We interpret $X$ as the type of \emph{observations} that can be made by $\G$ that inform the choice of a strategy and hence action, and $Y$ as the type of \emph{moves} or \emph{choices}. By that a game $\G$ is a \emph{process} that maps observations $X$ to choices $Y$.
The types $R$ and $S$, on the other hand, are `dual' or `contravariant' types and this is indicated above by the arrows flowing in the reverse direction. We think of $R$ as the type of utility (type of values about which the players in $\G$ have preferences) that actions might generate. Thus utility functions that arise in standard game theory are simply functions $Y \rightarrow R$. The type $S$ is dually called the coutility since it represents values that are returned to the calling environment by the game so that they can become utility for other processes. This is seen clearly and formally within the definition of composition of open games where the utility of one game (acting as the environment for the other) is computed via the coplay function of the other game. We point this out explicitly when discussing the composition of games.

The most straightforward parts of the definition of open games are the first two components. It is intuitive that a game has a set of strategy profiles and that, given a strategy profile and an observation, we can run the strategy profile on the observation to obtain a choice. To give a simple concrete example, suppose $Y = A \times B$, and define
\[ \Sigma = (X \to A) \times (X \times A \to B) \]
and
\[ \P ((\sigma_1, \sigma_2), x) = (\sigma_1 (x), \sigma_2 (x, \sigma_1 (x))) \]
This represents a two-player game of \emph{perfect information}: the value $x$ is the input which the first player observes and then chooses $a$ using the strategy $\sigma_1$. The second player observes both $x$ and $a$ and chooses $b$ using $\sigma_2$. The pair $(a, b)$ is taken to be the output.

To gain an intuition of coplay, let us first consider a very simple situation.
You receive your monthly income and upon observing your bonus, you decide to buy a bottle of champagne. You
go to a wine shop and for the given price, you buy a bottle, which gives
you a certain utility. Here, the connection between choice and reaction, between what
you do and what comes back to you, is rather immediate. Deciding is
simple. 

\emph{Open} games model situations where the connection
between your action and what comes back to you is left open - in the same way as
with selection functions: by allowing for \emph{all} possible contexts. The only
prerequisite for contexts is being well-typed. 
Coming back to the example, suppose, instead of buying the bottle yourself, you
give an amount of money to your friend, a champagne aficionado, with the
request to buy a bottle for you. So, your action gives him some money, and you
expect a bottle of champagne and by that some utility back from him. He observes
your bonus and the amount of money given to him, and will take some action that will
bring back a bottle of champagne from a yet unspecified environment. Ultimately, he
will hand back a bottle of champagne which will create utility for you. This
handing back of utility is computed through his coplay function which computes the coutility to hand back. 

Note, in the example we used the concept of \emph{utility}, which, by
definition, is the real number maximised by classical game theoretic agents. There is no
requirement in the definition of an open game that outcomes are either real
numbers, or a linear resource such as money nor that some utility is maximised at all. 
Indeed, slightly changing the example, we could
have defined coutility as belonging to a certain set of possible types of wine.
This flexibility may be seen as a necessary side-effect of allowing compositionality, but can itself be useful in modelling \cite{hedges_etal_higher_order_decision_theory,hedges_etal_selection_equilibria_higher_order_games}.

Possibly the most important part of the definition is the best response
relation, which is defined relative to an arbitrary \emph{context} -- as in the
case of games modelled via selection functions.
Hence, an open game also has Nash equilibria relative to an arbitrary context. A
context consists of a \emph{state}, which says what happened in the past, and a utility function, expressed 
as a \emph{continuation}, which says what will happen in the future. Compositional game theory relies on the observation that such \emph{relative} best response relations provide a strategic representation of games that can be composed.

{\em Relationship with Lenses:} Pairs of functions $X \to Y$ and $X \times R \to S$ are equivalent to \emph{polymorphic lenses} \cite{pickering_gibbons_wu_profunctor_optics,hedges_coherence_lenses_open_games}. Moreover, all open games can be built from lenses in a way we now describe. Note first that 
\begin{itemize}
\item An element $x : X$ is given by a lens $(1,1) \rightarrow (X,S)$. Call such lenses points $\operatorname{Pt}(X,S)$
\item A function $Y \rightarrow R$ is a lens $(Y,R) \rightarrow (1,1)$. Call such lenses co-points $\operatorname{CoPt}(Y,R)$
\end{itemize}

\begin{lemma}
An open game $\G : (X, S) \to (Y, R)$ is exactly 
\begin{itemize}
\item A family of lenses, that is a set $\Sigma_\G$  and, for each $\sigma : \Sigma_\G$, a lens $\G_{\sigma}: (X,S) \rightarrow (Y,R)$.
\item $\B_\G : \operatorname{Pt}(X,S) \times \operatorname{CoPt}(Y,R) \to \operatorname{Rel} (\Sigma_\G)$ 
\end{itemize}
\end{lemma}
In such a situation, we often save notational overhead by writing the lens $\G_{\sigma}$ as $\sigma$ in the following way
\[
(X,S) \stackrel{\sigma}{\longrightarrow}(Y,R)
\]

Lenses play a key role in the development of open games as they hide many details which otherwise severely pollute the presentation. More concretely, they encapsulate the purely algebraic parts of open games, leaving us to focus on the strategic behaviour of the best-response function. They also ensure all reasoning about open games can take place diagrammatically in the category of lenses. As a result, more recent work on open games has heavily exploited this connection \cite{hedges_morphisms_open_games}.

{\em Relationship with Geometry of Interaction:}  It has been pointed out to us that there is a connection between the geometry of interaction (GoI) and open games. A central construction within GoI, the Int-construction takes a traced monoidal category $\C$ and constructs another category $I(\C)$ whose objects are pairs of objects of $\C$, and whose morphisms $(X,S) \rightarrow (Y,R)$ are maps $X\otimes R \rightarrow Y \otimes S$. If $\C$ is cartesian, this is equivalent to  two functions $X \times R \rightarrow Y$ and $X \times R \rightarrow S$. If the former function does not use the input $R$, we get precisely a lens. This restriction means that open games are not completely symmetric and this explains why a trace operator is not needed to define composition --- one can simply calculate the forwards/covariant part of the composition and use that to calculate the backwards/contravariant part. 

When the base category is traced cartesian monoidal, both
lenses and the Int-construction are defined, and the two composition operators
agree. This means that there is an identity-on-objects functor from the category
of lenses to $I (\mathcal C)$, and this suggests it will be possible to build a more symmetrical theory of open games with \emph{computable} strategies.

\section{Examples of open games}\label{sec:examples}

An open game $\G : (1, 1) \to (1, 1)$, where $1 = \{ * \}$, consists (up to isomorphism of sets) of a set $\Sigma_\G$ of strategy profiles and a best response relation $\B_\G : \Rel (\Sigma_\G)$.
Since $(1, 1)$ will turn out to be the monoidal unit of the monoidal category of open games, we call $\G$ a \emph{scalar open game}, following the terminology of \cite{abramsky05}.

Any existing notion of `game' from which we can define a best response function can be encoded as a scalar open game.
For example, consider the prisoner's dilemma from Section \ref{sec:games}.
This can be represented as a scalar open game $\G : (1, 1) \to (1, 1)$ with
$\Sigma_\G = \{ C, D \}^2$ with the constant best response relation $\B_\G$
defined by $(\sigma, \sigma') \in \B_\G$ iff $\sigma' = (D, D)$. 

In this paper we will show that open games can be used to build such examples compositionally.
In the remainder of this section we will define families of open games that we consider \emph{atomic}, in the sense that they are not built compositionally from smaller games.
These families are \emph{decisions}, \emph{functions}, and \emph{counits}.

A decision is an open game that represents a single choice made by an agent.

\begin{definition}\label{def:maximising-decision}
	Let $X$ and $Y$ be sets.
	A (utility-maximising) \emph{decision} $\D : (X, 1) \to (Y, \RR)$ is an open game defined by the following data:
	\begin{itemize}
		\item $\Sigma_\D = X \to Y$
		\item $\P_\D (\sigma, x) = \sigma (x)$
		\item $\C_\D (\sigma, x, r) = *$
		\item $(\sigma, \sigma') \in \B_\D (x, k)$ iff $\sigma' (x) \in \argmax k$
	\end{itemize}
\end{definition}

A (pure) strategy for a single decision is a function that maps possible observations that can be made by the agent, to possible choices.
Such a strategy is considered optimal in the context $(x, k)$ iff it maps the current state $x$ to a maximising point of the current continuation $k$.

The reason that the pre-deviation strategy $\sigma$ plays no role in $(\sigma, \sigma') \in \B_\D (x, k)$ is that $\sigma$ is considered the strategy played by all \emph{other} players besides the one currently under consideration, and so it plays no role in a 1-player open game such as $\D$.
Put it in another way: in a one-player game there is nothing for $\sigma'$ to be a best response to.

The definition of a decision assumes an agent who maximises real-valued utility.
However, there is nothing inherent in the definition of an open game that says
this must be the case. Indeed, one can argue that the additional generality is
necessary to be compositional: An aggregate of two maximising agents, if
modelled as a selection function-like object, is not necessarily maximising. The
prisoner's dilemma is a standard example of an aggregate with behaviour that is
not (globally) maximising, that is to say, if the two players were modelled as
a single entity, they would choose  $(C,C)$. This is the sense in which selection functions are a theoretical precursor to open games.

\begin{definition}
	Let $X$, $Y$ and $R$ be sets, and let $\delta : (Y \to R) \to \mathcal P (Y)$ be a multi-valued selection function.
	We define an open game $\D_\delta : (X, 1) \to (Y, R)$ by the following data:
	\begin{itemize}
		\item $\Sigma_{\D_\delta} = X \to Y$
		\item $\P_{\D_\delta} (\sigma, x) = \sigma (x)$
		\item $\C_{\D_\delta} (\sigma, x, r) = *$
		\item $(\sigma, \sigma') \in \B_{\D_\delta} (x, k)$ iff $\sigma' (x) \in \delta (k)$
	\end{itemize}
\end{definition}

We will give an example of an open game with non-utility-maximising decisions in Section \ref{sec:simultaneous}.

Decisions are the only atomic open games that play a role in strategic reasoning.
We formalise this with the following definition.

\begin{definition}
	Let $\G : (X, S) \to (Y, R)$ be an open game.
	We call $\G$ \emph{strategically trivial} if it satisfies the following two conditions:
	\begin{itemize}
		\item $|\Sigma_\G| = 1$ (say, $\Sigma_\G = \{ * \}$)
		\item $(*, *) \in \B_\G (x, k)$ for all contexts $(x, k)$
	\end{itemize}
\end{definition}

The first condition says that there is exactly one strategy, and so there is no choice to be made.
The second condition says that this trivial strategy can never fail to be in equilibrium.
The idea behind this is that if a strategy profile fails to be an equilibrium,
it should always be because some player has an incentive to deviate.
Strategically trivial open games could also be called zero-player open games.

\begin{definition}
	Let $f : X \to Y$ and $g : R \to S$ be functions.
	We define a strategically trivial open game $(f, g) : (X, S) \to (Y, R)$ with play function $\P_{(f, g)} (x) = f (x)$ and coplay function $\C_{(f, g)} (*, x, r) = g (r)$.
\end{definition}

As a special case of this, a function $f : X \to Y$ can be `lifted' to an open game in two ways: covariantly as $(f, \id_1) : (X, 1) \to (Y, 1)$, or contravariantly as $(\id_1, f) : (1, Y) \to (1, X)$. 

The final class of atomic open games are the \emph{counits}.

\begin{definition}
	Let $X$ be a set.
	We define a strategically trivial open game $\varepsilon_X : (X, X) \to (1, 1)$ called a \emph{counit}, with play function $\P_{\varepsilon_X} (*, x) = *$ and coplay function $\C_{\varepsilon_X} (*, x, *) = x$.
\end{definition}

Recall that backward-flowing values are `teleological', that is, they are future values about which agents are reasoning.
The role of counits is to identify an ordinary forward-flowing value as the value \emph{about which} some past agent is reasoning. 
This plays an important role in the diagrammatic language of open games in the next section. Note there is no dual unit of type $(1, 1) \to (X, X)$ (the reader might like to try to define one) which is a reflection of the fact that game theory is not symmetric in its forward and backward looking facets. Mathematically, open games will not form a compact closed category.

\section{The monoidal category of open games}\label{sec:category}

In this section we will define a pair of operators for composing open games, categorical composition and monoidal product, which correspond to  \emph{sequential play} and \emph{simultaneous play}.
These two operators make open games into the morphisms of a symmetric monoidal category (after quotienting by isomorphisms of strategy sets).
As morphisms of a monoidal category open games can also be denoted by string
diagrams, and we introduce this diagrammatic language in the next section. While
these two operators form the core, we do not claim that they are a \emph{complete} basis of operators for building open games in any formal or informal sense.
Indeed, other operators are discussed in
\cite{ghani_kupke_lambert_forsberg_compositional_treatment_iterated_open_games}.

First we give the definition of categorical composition.
This is a form of sequential composition in which the choice made by the first component is \emph{hidden}, visible only to the second component but not to the outside.
Sequential play is more intuitive when the choices made by both components are visible; this will be recovered as a derived operator in Section \ref{sec:sequential}.

\begin{definition}
	Given a pair of open games $\G : (X, S) \to (Y, R)$ and $\H : (Y, R) \to (Z, Q)$, we define their composition $\H \circ \G : (X, S) \to (Z, Q)$ as follows.
	The set of strategy profiles is the cartesian product
	\[ \Sigma_{\H \circ \G} = \Sigma_\G \times \Sigma_\H \]
	The play function composes simply by composition of functions:
	\[ \P_{\H \circ \G} ((\sigma, \tau), x) = \P_\H (\tau, \P_\G (\sigma, x)) \]
	The coplay function composes as follows:
	\[ \C_{\H \circ \G} ((\sigma, \tau), x, q) = \C_\G (\sigma, x, \C_\H (\tau, \P_\G (\sigma, x), q)) \]
	The best response relation
	\[ ((\sigma, \tau), (\sigma', \tau')) \in \B_{\H \circ \G} (x, k) \]
	holds iff
	\[ (\sigma, \sigma') \in \B_\G (x, k') \]
	and
	\[ (\tau, \tau') \in \B_\H (\P_\G (\sigma, x), k) \]
	where $k' : Y \to R$ is defined by
	\[ k' (y) = \C_\H (\tau, y, k (\P_\H (\tau, y))) \]
\end{definition}

	Since the set of strategy profiles is usually a tuple consisting of a strategy for each decision, the condition $\Sigma_{\H \circ \G} = \Sigma_\G \times \Sigma_\H$ corresponds roughly to saying that the set of decisions in $\H \circ \G$ is the disjoint union of the decisions in $\G$ and $\H$.
	The play function says that to play the sequential composition $\H \circ \G$ in state $x$ with strategy profile $(\sigma, \tau)$, we first play $\G$ with $\sigma$ in state $x$, obtaining a state $y$ for $\H$, which we then play with $\tau$.
	
	The formula for composing coplay functions is hard to understand intuitively, but successfully captures the informal intuition given for coplay in section \ref{sec:open-games}.
	Alternatively, it can be seen as composition of lenses \cite{hedges_coherence_lenses_open_games}.
	
	Finally, we give the conditions for a strategy profile $(\sigma', \tau')$ to be a best response to $(\sigma, \tau)$ in a context $(x, k)$.
	This means that for each player in $\G$, $\sigma'$ must be rational assuming that all other players in $\G$ play $\sigma$ and all players in $\H$ play $\tau$, and also that for each player in $\H$, $\tau'$ must be rational assuming that all players in $\G$ play $\sigma$ and all other players in $\H$ play $\tau$.
	We can apply the compositionally-known best response relations for $\G$ and $\H$, after using these assumptions to appropriately modify the context.
	For $\H$ the continuation remains $k$, and the state $x$ is modified to $\P_\G (\sigma, x)$ using the assumption that players in $\G$ play $\sigma$.
	For $\G$ the state remains $x$, and the continuation is modified to $k'$, using the interpretation of the coplay function $\C_\H$ as the utility passed backward from $\H$ to $\G$, with the assumption that players in $\H$ play $\tau$.
	
Open games trivially cannot form a category, because this composition operator is not associative on the nose: the strategy profiles are
\begin{align*}
	\Sigma_{\mathcal I \circ (\H \circ \G)} &= (\Sigma_\G \times \Sigma_\H) \times \Sigma_\mathcal I \\
	&\neq \Sigma_\G \times (\Sigma_\H \times \Sigma_\mathcal I) \\
	&= \Sigma_{(\mathcal I \circ \H) \circ \G}
\end{align*}
There are three approaches to this problem.
The first, which is perfectly successful in practice, is to simply ignore it and informally work up to isomorphic strategy sets.
The second, which is attractive from a theoretical point of view, is to define a bicategory of open games in which the 2-cells are functions between strategy sets that suitably commute with the remaining structure.
The reason we do not take this approach is that monoidal bicategories are notoriously complicated, and generalising to a monoidal double category (whose axioms are typically much easier to verify \cite{shulman10}) would take us too far afield.
(This is carried out in \cite{hedges_morphisms_open_games}.)
Therefore in this paper we take the third approach, which is to define a
suitable equivalence relation on open games and then a category whose morphisms are equivalence classes.
This is equivalent to first defining a bicategory, and then obtaining a 1-category by quotienting by invertible 2-cells.

\begin{definition}
	Let $\G_1, \G_2 : (X, S) \to (Y, R)$ be open games.
	We write $\G_1 \sim \G_2$ if there is a bijection $i : \Sigma_{\G_1} \to \Sigma_{\G_2}$ such that
	\begin{itemize}
		\item $\P_{\G_1} (\sigma, x) = \P_{\G_2} (i (\sigma), x)$ for all $x : X$ and $\sigma : \Sigma_{\G_1}$
		\item $\C_{\G_1} (\sigma, x, r) = \C_{\G_2} (i (\sigma), x, r)$ for all $x : X$, $r : R$ and $\sigma : \Sigma_{\G_1}$
		\item $(\sigma, \sigma') \in \B_{\G_1} (x, k)$ iff $(i (\sigma), i (\sigma')) \in \B_{\G_2} (x, k)$ for all $x : X$, $k : Y \to R$ and $\sigma, \sigma' : \Sigma_{\G_1}$
	\end{itemize}
\end{definition}

\begin{proposition}
	For each type $(X, S) \to (Y, R)$, $\sim$ is an equivalence relation on the class of open games of that type.
\end{proposition}

The quotient under $\sim$ identifies open games with isomorphic strategy profiles and best responses.
This is close in spirit to the concept of \emph{best response equivalence}
of games in classical game theory \cite[p. 52f]{Myerson1991game}, but has some strange consequences.
For example, let $\G$ be the scalar open game representing the prisoner's dilemma from the previous section, with $\Sigma_\G = \{ C, D \}^2$ and $(\sigma, \sigma') \in \B_\G$ iff $\sigma' = (D, D)$.
Now consider a 1-player game with 4 choices $A = \{ 1, 2, 3, 4 \}$ and utility function $k : A \to \RR$ given by $k (x) = x$.
Encoding this as a scalar open game $\H$ yields $\Sigma_\H = A$ and $(\sigma, \sigma') \in \B_\H$ iff $x' = 4$, since $4$ is again a dominant strategy for the player.
Then $\G \sim \H$, and so in the quotient they will be \emph{equal} elements of the monoid of scalars, despite the fact that they represent games with different numbers of players.
However, this is a technical consequence of working with an equivalence relation, and does not happen if we instead use a bicategory or double category of open games.

\begin{proposition}
	$\circ$ is well-defined on equivalence classes, that is to say, if $\G \sim \G'$ and $\H \sim \H'$ then $\H \circ \G \sim \H' \circ \G'$.
\end{proposition}

\begin{definition}
	For each object $(X, S)$, the identity open game $\id_{(X, S)} : (X, S) \to (X, S)$ is the strategically trivial open game with
	\begin{itemize}
		\item $\P_{\id_{(X, S)}} (*, x) = x$
		\item $\C_{\id_{(X, S)}} (*, x, s) = s$
	\end{itemize}
\end{definition}

\begin{proposition}
	There is a category $\Game$ whose objects are pairs of sets and whose morphisms are equivalence classes of open games.
	The composition is $\circ$ and the identity on $(X, S)$ is the equivalence class of $\id_{(X, S)}$.
\end{proposition}

In the previous section we defined a strategically trivial open game $(f, g) : (X, S) \to (Y, R)$ given functions $f : X \to Y$ and $g : R \to S$.
This defines a functor $(-, -) : \Set \times \Set^\op \to \Game$.

Next we define the monoidal product of open games, which corresponds to simultaneous play.

\begin{definition}
	Let $\G_1 : (X_1, S_1) \to (Y_1, R_1)$ and $\G_2 : (X_2, S_2) \to (Y_2, R_2)$ be open games.
	We define an open game $\G_1 \otimes \G_2 : (X_1 \times X_2, S_1 \times S_2) \to (Y_1 \times Y_2, R_1 \times R_2)$ as follows:
	\begin{itemize}
		\item The set of strategy profiles is $\Sigma_{\G_1 \otimes \G_2} = \Sigma_{\G_1} \times \Sigma_{\G_2}$
		\item The play function is
		\[ \P_{\G_1 \otimes \G_2} ((\sigma_1, \sigma_2), (x_1, x_2)) = (\P_{\G_1} (\sigma_1, x_1), \P_{\G_2} (\sigma_2, x_2)) \]
		\item The coplay function is
		\begin{align*}
			&\C_{\G_1 \otimes \G_2} ((\sigma_1, \sigma_2), (x_1, x_2), (r_1, r_2)) \\
			=\ &(\C_{\G_1} (\sigma_1, x_1, r_1), \C_{\G_2} (\sigma_2, x_2, r_2))
		\end{align*}
		\item The relation
		\[ ((\sigma_1, \sigma_2), (\sigma'_1, \sigma'_2)) \in \B_{\G_1 \otimes \G_2} ((x_1, x_2), k) \]
		holds iff the relations $(\sigma_1, \sigma'_1) \in \B_{\G_1} (x_1, k_1)$ and $(\sigma_2, \sigma'_2) \in \B_{\G_2} (x_2, k_2)$ both hold, where
		\[ k_1 : Y_1 \to R_1 \qquad k_2 : Y_2 \to R_2 \]
		are defined by
		\[ k_1 (y_1) = \pi_1 (k (y_1, \P_{\G_2} (\sigma_2, x_2))) \]
		\[ k_2 (y_2) = \pi_2 (k (\P_{\G_1} (\sigma_1, x_1), y_2)) \]
	\end{itemize}
\end{definition}

\begin{lemma}
	$\otimes$ is well-defined on equivalence classes.
\end{lemma}

\begin{lemma}
	$\otimes$ defines a bifunctor $\Game \times \Game \to \Game$.
\end{lemma}

\begin{theorem}
	$\Game$ is a monoidal category in which the monoidal product on objects is $(X_1, S_1) \otimes (X_2, S_2) = (X_1 \times X_2, S_1 \times S_2)$ and that on morphisms is previously defined.
	The monoidal unit is the object $I = (1, 1)$.
\end{theorem}

The structure morphisms of the monoidal category are inherited from the monoidal category $\Set \times \Set^\op$ via the functor $(-, -)$, where $\Set$ is cartesian monoidal.
$\Game$ is moreover symmetric monoidal, with the symmetry inherited from $\Set \times \Set^\op$.

{\em A Lens-theoretic View:} A cleaner approach arises if one factors the definition of parallel and sequential composition via the use of lenses. First note that the above definitions restrict to lenses meaning that the category of lenses is symmetric monoidal. Just as open games can be defined in terms of lenses, the categorical structure of open games can similarly be defined in terms of the simpler categorical structure of lenses.

\begin{lemma}
	Let $\G:(X,S) \rightarrow (Y,R)$ and $\H:(Y,R) \rightarrow (Z,T)$ be open games.  Then the composite $\H \circ \G$ is the family of lenses indexed by $\Sigma_G \times \Sigma_H$ with the pair $(\sigma,\tau)$ indexing the lens
	\[
	(X,S) \stackrel{\sigma}{\longrightarrow} (Y,R) \stackrel{\tau}{\longrightarrow} (Z,T) 
	\]
	Given a point $x:(1,1) \rightarrow (X,S)$ and a copoint $x:(Z,T) \rightarrow (1,1)$, then $((\sigma,\tau), (\sigma',\tau')) \in \B_{\H\circ \G} (x,k)$
	holds iff
	\begin{itemize}
	\item $(\sigma,\sigma') \in \B_{\G} (x, k \circ \tau)$, and
	\item $(\tau,\tau') \in \B_{\H} (\sigma \circ  x, k)$
	\end{itemize}
	\end{lemma}
Notice how lens composition hides all the details of how play and coplay functions knit together in the composite game. In particular, in the definition of composition, the function $k'(y) = \C_{\H}(\tau, y, k(\P_{\H}(\tau, y)))$ is merely the lens composite $k \circ \tau$.  A similarly simplified construction of the monoidal product of open games via the monoidal product of lenses can also be given.

\section{String diagrams}\label{sec:string_diagram}

We will now informally present the string diagram language for open games.
A formal presentation can be found in \cite{hedges_coherence_lenses_open_games}, which proves a coherence theorem by which we can \emph{define} an open game by its string diagram, given interpretations of the atomic open games.
We refer the reader to \cite{selinger11} for a summary of graphical languages of this sort.
The language of open games is an extension of those of symmetric monoidal categories, and similar to (but not exactly) a fragment of compact closed categories.

The key idea of the string diagram language is to treat the object $(X, S)$ as a formal tensor product $X \otimes S^*$, where $-^*$ is a duality that is defined on objects but not on arbitrary open games.
Diagrammatically we represent this duality by an orientation on strings, so a general object $(X, S)$ is denoted by
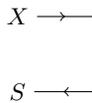
\begin{center} \begin{tikzpicture}
	\node (X1) at (0, 1) {$X$}; \node (S2) at (0, 0) {$S$};
	\draw [->-] (X1) to (1, 1); \draw [->-] (1, 0) to (S2);
\end{tikzpicture} \end{center}
and thus a general open game $\G : (X, S) \to (Y, R)$ is denoted by
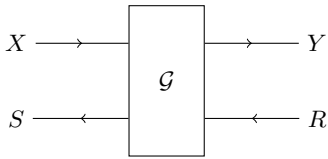
\begin{center} \begin{tikzpicture}
	\node (X) at (-2, .5) {$X$}; \node (Y) at (2, .5) {$Y$}; \node (R) at (2, -.5) {$R$}; \node (S) at (-2, -.5) {$S$};
	\node [rectangle, minimum height=2cm, minimum width=1cm, draw] (G) at (0, 0) {$\G$};
	\draw [->-] (X) to (G.west |- X); \draw [->-] (G.east |- Y) to (Y); \draw [->-] (R) to (G.east |- R); \draw [->-] (G.west |- S) to (S);
\end{tikzpicture} \end{center}

More formally, we allow individual strings to represent \emph{covariant objects} of the form $(X, 1)$, and \emph{contravariant objects} of the form $(1, S)$.
Then, up to isomorphism, a general object can be written as a tensor product $(X, S) \cong (X, 1) \otimes (1, S)$ of a covariant and a contravariant object.
If we define a duality operation on arbitrary objects by $(X, S)^* = (S, X)$, then $(X, S) \cong (X, 1) \otimes (S, 1)^*$.
This justifies the informal statement $(X, S) = X \otimes S^*$, because we can identify $\Set$ with a symmetric monoidal subcategory $\Set \hookrightarrow \Game$ by identifying $X$ with $(X, 1)$ and $f$ with $(f, \id_1)$.

Notice that since we also have $(X, S) \cong (S, 1)^* \otimes (X, 1)$, the open game $\G$ can equally be denoted
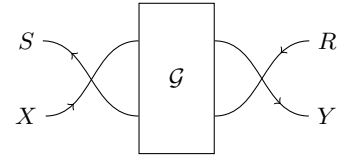
\begin{center} \begin{tikzpicture}
	\node (X) at (-2, -.5) {$X$}; \node (Y) at (2, -.5) {$Y$}; \node (R) at (2, .5) {$R$}; \node (S) at (-2, .5) {$S$};
	\node [rectangle, minimum height=2cm, minimum width=1cm, draw] (G) at (0, 0) {$\G$};
	\draw [decoration={markings, mark=at position .25 with {\arrow{>}}}, postaction={decorate}] (X) to [out=0, in=180] (G.west |- S);
	\draw [decoration={markings, mark=at position .75 with {\arrow{>}}}, postaction={decorate}] (G.east |- R) to [out=0, in=180] (Y);
	\draw [decoration={markings, mark=at position .25 with {\arrow{>}}}, postaction={decorate}] (R) to [out=180, in=0] (G.east |- Y);
	\draw [decoration={markings, mark=at position .75 with {\arrow{>}}}, postaction={decorate}] (G.west |- X) to [out=180, in=0] (S);
\end{tikzpicture} \end{center}
That is, the relative ordering of covariant and contravariant parts of an object does not matter.
More formally, the objects $X \otimes S^*$ and $S^* \otimes X$ are equal in the strictification of $\Game$, and the symmetry $s_{X, S^*}$ is an identity.

Corresponding to each atomic open game we have a corresponding `atomic' string diagram, which we compose by the usual operations of end-to-end and side-by-side juxtaposition.
For example, a utility-maximising decision $\D : (X, 1) \to (Y, \RR)$ corresponds to a node
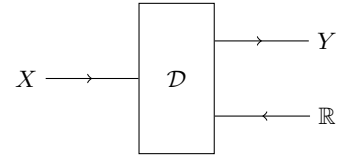
\begin{center} \begin{tikzpicture}
	\node (X) at (-2, 0) {$X$}; \node (Y) at (2, .5) {$Y$}; \node (R) at (2, -.5) {$\RR$};
	\node [rectangle, minimum height=2cm, minimum width=1cm, draw] (G) at (0, 0) {$\D$};
	\draw [->-] (X) to (G); \draw [->-] (G.east |- Y) to (Y); \draw [->-] (R) to (G.east |- R);
\end{tikzpicture} \end{center}
If $X$ is a 1-element set, we further restrict this to
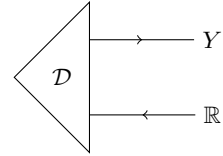
\begin{center} \begin{tikzpicture}
	\node (Y2) at (8, .5) {$Y$}; \node (R2) at (8, -.5) {$\mathbb R$};
	\node [isosceles triangle, isosceles triangle apex angle=90, shape border rotate=180, minimum width=2cm, draw] (D2) at (6, 0) {$\D$};
	\draw [->-] (D2.east |- Y2) to (Y2); \draw [->-] (R2) to (D2.east |- R2);
\end{tikzpicture} \end{center}
with the usual (purely syntactic) convention of using a triangle for morphisms into or out of the monoidal unit.

Given a function $f : X \to Y$, its covariant lifting $(f, \id_1) : (X, 1) \to (Y, 1)$ and its contravariant lifting $(\id_1, f) : (1, Y) \to (1, X)$ are respectively denoted
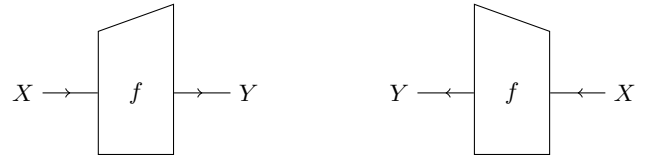
\begin{center} \begin{tikzpicture}
	\node (X1) at (0, 0) {$X$}; \node (Y1) at (3, 0) {$Y$};
	\node [trapezium, trapezium left angle=0, trapezium right angle=75, shape border rotate=90, trapezium stretches=true, minimum height=1cm, minimum width=2cm, draw] (f1) at (1.5, 0) {$f$};
	\draw [->-] (X1) to (f1); \draw [->-] (f1) to (Y1);
	\node (X2) at (8, 0) {$X$}; \node (Y2) at (5, 0) {$Y$};
	\node [trapezium, trapezium left angle=75, trapezium right angle=0, shape border rotate=270, trapezium stretches=true, minimum height=1cm, minimum width=2cm, draw] (f2) at (6.5, 0) {$f$};
	\draw [->-] (X2) to (f2); \draw [->-] (f2) to (Y2);
\end{tikzpicture} \end{center}
The syntax of trapezia under reflection (with unoriented strings) is used for
the adjoint of a linear map by Bob Coecke and others, for example in \cite{coecke_kissinger_picturing_quantum_processes}.

The deleting function $X \to 1$ and diagonal function $X \to X^2$ lift to give a commutative comonoid on every covariant object, and a commutative monoid on every contravariant object.
We give these the special syntax
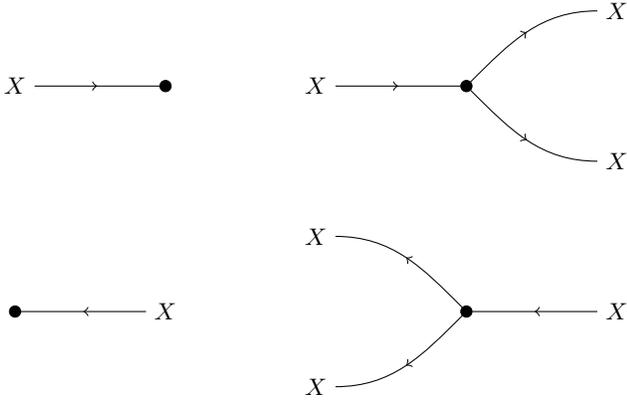
\begin{center} \begin{tikzpicture}
	\node (X1) at (0, 1) {$X$}; \node [circle, scale=0.5, fill=black, draw] (o) at (2, 1) {};
	\draw [->-] (X1) to (o);
	\node (X2) at (4, 1) {$X$}; \node [circle, scale=0.5, fill=black, draw] (m) at (6, 1) {};
	\node (X3) at (8, 2) {$X$}; \node (X4) at (8, 0) {$X$};
	\draw [->-] (X2) to (m);
	\draw [->-] (m) to [out=45, in=180] (X3); \draw [->-] (m) to [out=-45, in=180] (X4);
	\node [circle, scale=0.5, fill=black, draw] (o) at (0, -2) {}; \node (X1) at (2, -2) {$X$};
	\draw [->-] (X1) to (o);
	\node (X2) at (4, -1) {$X$}; \node (X3) at (4, -3) {$X$};
	\node [circle, scale=0.5, fill=black, draw] (m) at (6, -2) {}; \node (X4) at (8, -2) {$X$};
	\draw [->-] (X4) to (m);
	\draw [->-] (m) to [out=135, in=0] (X2); \draw [->-] (m) to [out=-135, in=0] (X3);
\end{tikzpicture} \end{center}

The final atomic open games that we must give representations to are the counits $\varepsilon_X : (X, X) \to I$.
This is denoted by a bending wire
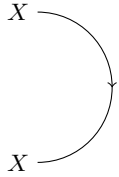
\begin{center} \begin{tikzpicture}
	\node (X1) at (0, 2) {$X$}; \node (X2) at (0, 0) {$X$};
	\draw [->-] (X1) to [out=0, in=90] (1.25, 1) to [out=-90, in=0] (X2);
\end{tikzpicture} \end{center}
Since there is no natural strategically trivial open game $I \to (X, X)$, we do not allow wires to bend in the opposite direction in our string diagrams.

Covariant functions, contravariant functions and counits are related by the \emph{counit law}, stating that the string diagrams
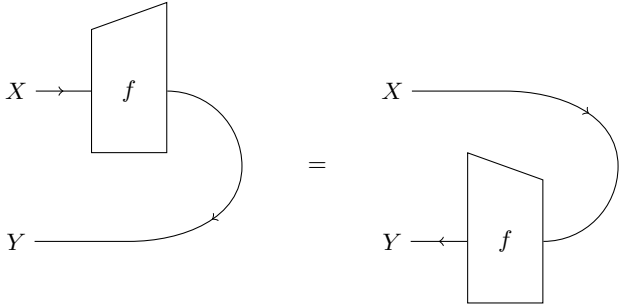
\begin{center} \begin{tikzpicture}
	\node (X1) at (0, 2) {$X$}; \node (Y1) at (0, 0) {$Y$};
	\node [trapezium, trapezium left angle=0, trapezium right angle=75, shape border rotate=90, trapezium stretches=true, minimum height=1cm, minimum width=2cm, draw] (f1) at (1.5, 2) {$f$};
	\draw [->-] (X1) to (f1); \draw [->-] (f1) to [out=0, in=90] (3, 1) to [out=-90, in=0] (1.5, 0) to (Y1);
	\node at (4, 1) {$=$};
	\node (X2) at (5, 2) {$X$}; \node (Y2) at (5, 0) {$Y$};
	\node [trapezium, trapezium left angle=75, trapezium right angle=0, shape border rotate=270, trapezium stretches=true, minimum height=1cm, minimum width=2cm, draw] (f2) at (6.5, 0) {$f$};
	\draw [->-] (X2) to (6.5, 2) to [out=0, in=90] (8, 1) to [out=-90, in=0] (f2); \draw [->-] (f2) to (Y2);
\end{tikzpicture} \end{center}
denote equal open games.
That is to say, the diagram of open games
\begin{center} \begin{tikzpicture}[node distance=4cm, auto]
	\node (A) {$(X, Y)$}; \node (B) [right of=A] {$(Y, Y)$};
	\node (C) [below of=A] {$(X, X)$}; \node (D) [below of=B] {$(1, 1)$};
	\draw [->] (A) to node {$(f, \id_1) \otimes \id_{(1, Y)}$} (B); \draw [->] (A) to node [left] {$\id_{(X, 1)} \otimes (\id_1, f)$} (C);
	\draw [->] (B) to node {$\varepsilon_Y$} (D); \draw [->] (C) to node {$\varepsilon_X$} (D);
\end{tikzpicture} \end{center}
commutes (where the monoidal structure morphisms are implicit).

Given a diagram built from the pieces we have described, which does not contain any wire bending in the illegal direction, we can compositionally build an open game where
\begin{itemize}
	\item Decision and function nodes are interpreted as the corresponding atomic open game
	\item Side-by-side and end-to-end composition of diagrams is interpreted as monoidal product and categorical composition of open games
	\item A backwards-bending wire is interpreted as the corresponding counit
\end{itemize}
The coherence theorem for teleological categories \cite{hedges_coherence_lenses_open_games} states that the resulting open game is invariant under topological manipulations of the diagram, including rotating function nodes around a bend using the counit law, provided that the new diagram does not contain a wire bending in the illegal direction.
Several examples of interpreting a diagram as an open game can be seen in the next two sections.

\section{Simultaneous move games}\label{sec:simultaneous}

In this section and the next we will apply the theory of the previous sections to demonstrate that various classes of games can be represented as open games and can be built compositionally.

We begin with simultaneous move games.
The decision $\D_{1, X_i} : I \to (X_i, \RR)$ represents an agent who makes a choice from a set $X_i$ in order to maximise a real number.

\begin{theorem}
	Let
	\[ \G := \bigotimes_{i = 1}^n \D_{1, X_i} : I \to \left( \prod_{i = 1}^n X_i, \RR^n \right) \]
	be a monoidal product of decisions.
	Then the set of strategy profiles of $\G$ is equal to the set of pure strategy
  profiles of a simultaneous move game with sets of choices $X_i$, namely
	\[ \Sigma_\G = \prod_{i = 1}^n X_i \]
	and, for any function $q : \prod_{i = 1}^n X_i \to \RR^n$, the relation $\B_\G
  (*, q) \subseteq \Sigma_\G \times \Sigma_\G$ is precisely the best response
  relation for the simultaneous move  game with outcome function $q$ (and, hence, the fixpoints of $\B_\G (*, q)$ are the pure strategy Nash equilibria).
\end{theorem}

In particular, by the associativity of the monoidal product, the monoidal product of an $m$-player and an $n$-player open game is an $m + n$-player open game.

Given a particular utility function $q : \prod_{i = 1}^n X_i \to \RR^n$, consider the covariant function
\[ (q, 1) : \left( \prod_{i = 1}^n X_i, 1 \right) \to (\RR^n, 1) \]
We take the monoidal product of this with an identity morphism and then post-compose with a counit, to yield the strategically trivial open game
\[ \left( \prod_{i = 1}^n X_i, \RR^n \right) \xrightarrow{(q, 1) \otimes \id_{(1, \RR^n)}} (\RR^n, \RR^n) \xrightarrow{\varepsilon_{\RR^n}} (1, 1) \]
By the counit law, this can be equivalently written as
\[ \varepsilon_{\prod_{i = 1}^n X_i} \circ (\id_{(\prod_{i = 1}^n X_i, 1)} \otimes (1, q)) \]
Now consider the scalar open game
\begin{align*}
	(1, 1) &\xrightarrow{\bigotimes_{i = 1}^n \D_{1, X_i}} \left( \prod_{i = 1}^n X_i, \RR^n \right) \\
	&\xrightarrow{(q, 1) \otimes \id_{(1, \RR^n)}} (\RR^n, \RR^n) \\
	&\xrightarrow{\varepsilon_{\RR^n}} (1, 1)
\end{align*}
This scalar open game has the property that its best response relation $\B (*,
*)$, for the unique context $(*, *)$, is precisely the best response function
for the simultaneous move game with outcome function $q$.

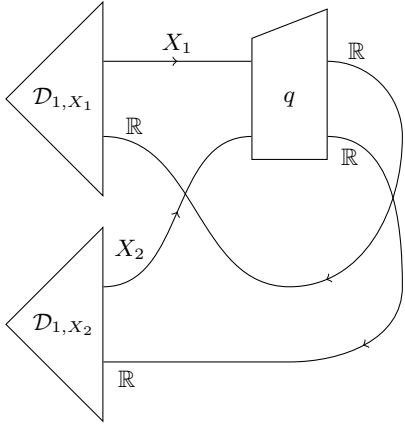
\begin{figure}
	\begin{center} \begin{tikzpicture}
		\node [isosceles triangle, isosceles triangle apex angle=90, shape border rotate=180, minimum width=2cm, draw] (D1) at (0, 3) {$\mathcal D_{1, X_1}$};
		\node [isosceles triangle, isosceles triangle apex angle=90, shape border rotate=180, minimum width=2cm, draw] (D2) at (0, 0) {$\mathcal D_{1, X_2}$};
		\node [trapezium, trapezium left angle=0, trapezium right angle=75, shape border rotate=90, trapezium stretches=true, minimum height=1cm, minimum width=2cm, draw] (U) at (3, 3) {$q$};
		\node (d1) at (0, -.5) {}; \node (d2) at (0, .5) {}; \node (d3) at (0, 2.5) {}; \node (d4) at (0, 3.5) {}; \node (d5) at (0, 1) {}; \node (d6) at (0, 2) {};
		\draw [->-] (D1.east |- d4) to node [above] {$X_1$} (U.west |- d4);
		\draw [->-] (D2.east |- d2) to [out=0, in=180] node [above=5pt, very near start] {$X_2$} (U.west |- d3);
		\draw [->-] (U.east |- d4) to [out=0, in=90] node [above, near start] {$\RR$} (4.5, 2.5) to [out=-90, in=0] (3, .5) to [out=180, in=0] node [above, very near end] {$\RR$} (D1.east |- d3);
		\draw [->-] (U.east |- d3) to [out=0, in=90] node [below, very near start] {$\RR$} (4.5, .5) to [out=-90, in=0] (3, -.5) to node [below, very near end] {$\RR$} (D2.east |- d1);
	\end{tikzpicture} \end{center}
	\caption{String diagram for simultaneous move game}
	\label{fig:simultaneous-game}
\end{figure}

For small values of $n$, we can visualise this scalar open game as a string diagram.
For example, when $n = 2$ the corresponding string diagram is depicted in Figure \ref{fig:simultaneous-game}.
Here, for the first time we can see how the information flow in a game is visualised with a string diagram: the utility generated by the utility function is `fed back' to each agent via a counit.

\begin{figure}
	\begin{center} \begin{tikzpicture}
		\node [isosceles triangle, isosceles triangle apex angle=90, shape border rotate=180, minimum width=2cm, draw] (D1) at (0, 3) {$\mathcal D_1$};
		\node [isosceles triangle, isosceles triangle apex angle=90, shape border rotate=180, minimum width=2cm, draw] (D2) at (0, 0) {$\mathcal D_2$};
		\node (d1) at (0, -.5) {}; \node (d2) at (0, .5) {}; \node (d3) at (0, 2.5) {}; \node (d4) at (0, 3.5) {};
		\draw [->-] (D1.east |- d4) to [out=0, in=90] node [above, very near start] {$X$} (3, 1.5) to [out=-90, in=0] node [below, very near end] {$X$} (D2.east |- d1);
		\draw [->-] (D2.east |- d2) to [out=0, in=-90] node [above, near start] {$X$} (2, 1.5) to [out=90, in=0] node [below, near end] {$X$} (D1.east |- d3);
	\end{tikzpicture} \end{center}
	\caption{Simultaneous move game with non-utility-maximising players}
	\label{fig:banana-game}
\end{figure}
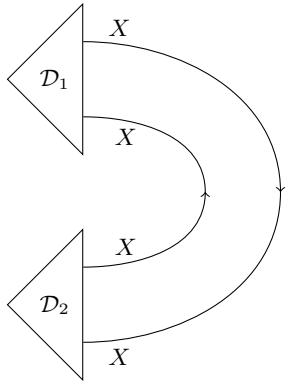

The previous results can be strengthened to an arbitrary higher order game (introduced in Section \ref{sec:games}) using the open games $\mathcal D_\delta$ associated to a selection function from Section \ref{sec:examples}, in which case the fixpoints of the best response relation are selection equilibria.
An interesting special case of this is depicted in Figure \ref{fig:banana-game}, in which the outcome that is `optimised' by each player is nothing but the choice of the other player.
(Ignoring the types, this string diagram arises from Figure
\ref{fig:simultaneous-game} by replacing $q$ with a symmetry, i.e. crossing
wires; this provides a nontrivial example of reasoning about the  equivalence of games from topological manipulations of string diagrams.)
Without imposing an order relation on the set of choices, it is not possible to interpret $\D_1$ and $\D_2$ as utility-maximising decisions.
Instead, by interpreting $\D_1$ and $\D_2$ in suitable ways we can obtain directly analogous results to the Keynes beauty contest example in \cite{hedges_etal_selection_equilibria_higher_order_games}.

We will consider three different ways of interpreting $\D_1$ and $\D_2$, which result in three different games.
If $\D_1 = \D_2$ are both $\D_\text{fix}$, where $\text{fix} : (X \to X) \to \mathcal P (X)$ is the selection function
\[ \text{fix} (k) = \{ x : X \mid x = k (x) \} \]
the resulting scalar open game is a coordination game.
In particular, if $X = \{ A, B \}$ then the best response relation is the same as \emph{Meeting in New York}, the 2-player simultaneous move game with utility maximising players and outcome function
\[ q (x, y) = \begin{cases}
	(1, 1) &\text{ if } x = y \\
	(0, 0) &\text{ if } x \neq y
\end{cases} \]
In particular, the pure Nash equilibria and the fixpoints of $\B_\G (*, *)$ are $(A, A)$ and $(B, B)$.

Next, we interpret both $\D_1$ and $\D_2$ as the open game lifted from the anti-fixpoint selection function
\[ \operatorname{anti-fix} (k) = \{ x \mid x \neq k (x) \} \]
If we do this, then the resulting scalar open game $\G$ has the
same best response relation as a simultaneous move game with 2 utility maximising players and outcome function
\[ q (x, y) = \begin{cases}
	(0, 0) &\text{ if } x = y \\
	(1, 1) &\text{ if } x \neq y
\end{cases} \]
This is an \emph{anti-coordination game}.
If $X = \{ A, B \}$ then the pure Nash equilibria and fixpoints of $\B_\G (*, *)$ are $(A, B)$ and $(B, A)$.

Finally, we interpret $\D_1 = \D_{\text{fix}}$ and $\D_2 = \D_{\operatorname{anti-fix}}$.
This is a game in which the first player would like to coordinate with the second, and the second would like to differentiate from the first.
This has the same best response relation as \emph{matching pennies}, the game with outcome function
\[ q (x, y) = \begin{cases}
	(1, 0) & \text{ if } x = y \\
	(0, 1) &\text{ if } x \neq y
\end{cases} \]
This game has no Nash equilibria in pure strategies.

\section{Sequential games}\label{sec:sequential}

In the previous section we showed that a simultaneous move game can be represented as an open game using monoidal products of decisions.
In this section we will represent sequential games, in which players can observe previous actions of other players before making their choice.

We focus on the sub-class of finite \emph{sequential games} from \cite{escardo11} introduced in Section \ref{sec:games}, which are the finite extensive-form games of perfect information in which at each stage a different player chooses, and the player choosing and the set of possible choices are determined only by the stage number and not the previous moves.
That is to say, distinct players $1, \ldots, n$ sequentially make choices from sets $X_1, \ldots, X_n$, with each player observing every previous move.
Relaxing each of these restrictions is possible but requires more work (generally, defining additional composition operators on open games, such as those in \cite{ghani_kupke_lambert_forsberg_compositional_treatment_iterated_open_games}), and so we focus on this sub-class for simplicity.

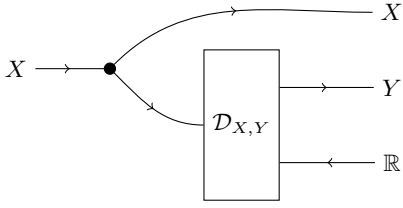
\begin{figure}
\begin{center} \begin{tikzpicture}
	\node (X) at (-3, .75) {$X$}; \node (Y) at (2, .5) {$Y$}; \node (R) at (2, -.5) {$\RR$}; \node (X2) at (2, 1.5) {$X$};
	\node [rectangle, minimum height=2cm, minimum width=1cm, draw] (G) at (0, 0) {$\D_{X, Y}$};
	\node [circle, scale=.5, fill=black, draw] (m) at (-1.75, .75) {};
	\draw [->-] (X) to (m); \draw [->-] (m) to [out=-45, in=180] (G); \draw [->-] (m) to [out=45, in=180] (X2);
	\draw [->-] (G.east |- Y) to (Y); \draw [->-] (R) to (G.east |- R);
\end{tikzpicture} \end{center}
\caption{Definition of $\D^\Delta_{X,Y}$};
\label{fig:definition-of-d-delta}
\end{figure}

Recall from Definition \ref{def:maximising-decision} that the choice of an element of $Y$ after observing an element of $X$ by a utility-maximising agent is modelled by the open game $\D_{X, Y} : (X, 1) \to (Y, \RR)$ defined by
\begin{itemize}
	\item $\Sigma_{\D_{X, Y}} = X \to Y$
	\item $\P_{\D_{X, Y}} (\sigma, x) = \sigma (x)$
	\item $\C_{\D_{X, Y}} (\sigma, x, r) = *$
	\item $(\sigma, \sigma') \in \B_{\D_{X, Y}} (x, k)$ iff $\sigma' (x) \in \argmax k$
\end{itemize}
The basic element of a sequential game is the open game $\D^\Delta_{X, Y} : (X, 1) \to (X \times Y, \RR)$ denoted by the string diagram in figure \ref{fig:definition-of-d-delta}.
Algebraically, this is $(\id_{(X, 1)} \otimes \D_{X, Y}) \circ (\Delta_X, 1)$, where $(\Delta_X, 1) : (X, 1) \to (X \times X, 1)$ is the lifting of the copying function $X \to X \times X$.
By applying the definitions of the composition operators $\circ$ and $\otimes$, the reader can verify that $\D^\Delta_{X, Y}$ is concretely given as follows, up to natural isomorphism:
\begin{itemize}
	\item $\Sigma_{\D^\Delta_{X, Y}} = X \to Y$
	\item $\P_{\D^\Delta_{X, Y}} (\sigma, x) = (x, \sigma (x))$
	\item $\C_{\D^\Delta_{X, Y}} (\sigma, x, r) = *$
	\item $(\sigma, \sigma') \in \B_{\D^\Delta_{X, Y}} (x, k)$, where $k : X \times Y \to \RR$, iff $\sigma' (x) \in \argmax_{y : Y} k (x, y)$
\end{itemize}

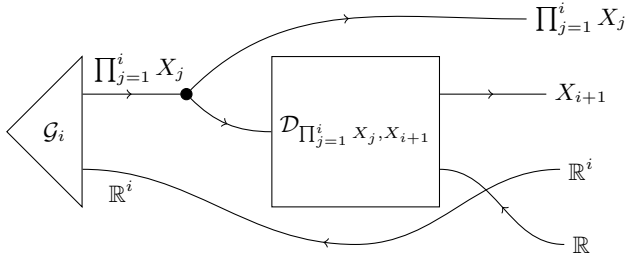
\begin{figure}
	\begin{center} \begin{tikzpicture}
		\node [isosceles triangle, isosceles triangle apex angle=90, shape border rotate=180, minimum width=2cm, draw] (G) at (0, 0) {$\G_i$};
		\node [circle, scale=.5, fill=black, draw] (m) at (1.75, .5) {};
		\node [rectangle, minimum height=2cm, minimum width=1cm, draw] (D) at (4, 0) {$\D_{\prod_{j = 1}^i X_j, X_{i + 1}}$};
		\node (X) at (7, 1.5) {$\prod_{j = 1}^i X_j$}; \node (Y) at (7, .5) {$X_{i + 1}$}; \node (R1) at (7, -.5) {$\RR^i$}; \node (R2) at (7, -1.5) {$\RR$};
		\draw [->-] (G.east |- m) to node [above] {$\ \ \prod_{j = 1}^i X_j$} (m); 
		\draw [->-] (m) to [out=45, in=180] (X);
		\draw [->-] (m) to [out=-45, in=180] (D);
		\draw [->-] (D.east |- m) to (Y);
		\draw [->-] (R1) to [out=180, in=0] (4, -1.5) to [out=180, in=0] node [below, very near end] {$\RR^i$} (G.east |- R1);
		\draw [->-] (R2) to [out=180, in=0] (D.east |- R1);
	\end{tikzpicture} \end{center}
	\caption{Recursive step of definition \ref{def:sequential-open-game}}
	\label{fig:sequential-game-recursive-step}
\end{figure}

\begin{definition}\label{def:sequential-open-game}
	Let $X_1, \ldots, X_n$ be a sequence of sets.
	We recursively define a sequence of open games
	\[ \G_i : (1, 1) \to \left( \prod_{j = 1}^i X_j, \RR^i \right) \]
	as follows.
	The base case is $\G_0 = \id_{(1, 1)} : (1, 1) \to (1, 1)$.
	In the recursive step, $\G_{i + 1}$ is defined in terms of $\G_i$ and $\D_{\prod_{j = 1}^i X_j, X_{i + 1}}$ by the string diagram in Figure \ref{fig:sequential-game-recursive-step}.
\end{definition}

\begin{theorem}
	Let $X_1, \ldots, X_n$ be a sequence of sets and $q : \prod_{i = 1}^n X_i \to \RR^n$.
	Let
	\[ \G_n : (1, 1) \to \left( \prod_{i = 1}^n X_i, \RR^n \right) \]
	be defined as in Definition \ref{def:sequential-open-game}.
	Then
	\[ \Sigma_{\G_n} = \prod_{i = 1}^n \left( \prod_{j = 1}^{i - 1} X_j \to X_i \right) \]
	is the set of strategy profiles of the $n$-player sequential game with outcome function $q$, and $\B_{\G_n} (*, q)$ is its best response relation.
\end{theorem}

\begin{figure}
	\begin{center} \begin{tikzpicture}
		\node [isosceles triangle, isosceles triangle apex angle=90, shape border rotate=180, minimum width=2cm, draw] (D1) at (0, 0) {$\D_{1, X}$};
		\node [circle, scale=.5, fill=black, draw] (m) at (1.75, .5) {};
		\node [rectangle, minimum height=2cm, minimum width=1cm, draw] (D2) at (3.5, 0) {$\D_{X, Y}$};
		\node [trapezium, trapezium left angle=0, trapezium right angle=75, shape border rotate=90, trapezium stretches=true, minimum height=1cm, minimum width=2cm, draw] (q) at (5.5, 1) {$q$};
		\node (d1) at (0, -.5) {}; \node (d2) at (0, 1.5) {};
		\draw [->-] (D1.east |- m) to node [above, near start] {$X$} (m);
		\draw [->-] (m) to [out=45, in=180] node [above, very near end] {$X$} (q.west |- d2);
		\draw [->-] (m) to [out=-45, in=180] node [below, near end] {$X$} (D2);
		\draw [->-] (D2.east |- m) to node [above] {$Y$} (q.west |-m);
		\draw [->-] (q.east |- d2) to [out=0, in=90] node [above, very near start] {$\RR$} (7, 0) to [out=-90, in=0] (3.5, -1.5) to [out=180, in=0] node [below, very near end] {$\RR$} (D1.east |- d1);
		\draw [->-] (q.east |- m) to [out=0, in=90] node [above, near start] {$\RR$} (6.5, 0) to [out=-90, in=0] node [above, very near end] {$\RR$} (D2.east |- d1);
	\end{tikzpicture} \end{center}
	\caption{Example sequential open game with 2 players}
	\label{fig:example-sequential-open-game}
\end{figure}
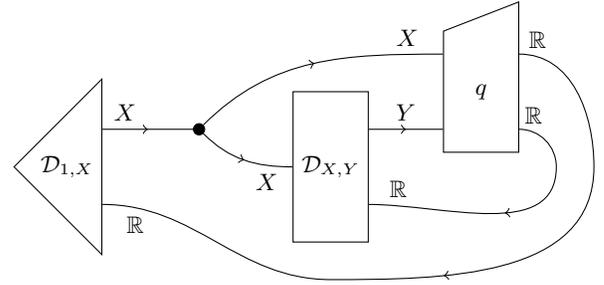

For example, a 2-player sequential game with outcome function $q : X \times Y \to \RR^2$ corresponds to the open game depicted in Figure \ref{fig:example-sequential-open-game}.
This game has a set of strategy profiles $\Sigma = X \times (X \to Y)$, and the best response relation $\B \subseteq \Sigma \times \Sigma$ is defined by $(\sigma, \sigma') \in \B$ iff
\[ \sigma'_1 \in \argmax_{x : X} (q (x, \sigma_2 (x)))_1 \]
and
\[ \sigma'_2 (\sigma_1) \in \argmax_{y : Y} (q (\sigma_1, y))_2 \]

\begin{figure}
	\begin{center} \begin{tikzpicture}
		\node [isosceles triangle, isosceles triangle apex angle=90, shape border rotate=180, minimum width=2cm, draw] (D1) at (0, 0) {$\D_{1, X}$};
		\node [circle, scale=.5, fill=black, draw] (m) at (1.25, .75) {};
		\node [circle, scale=.5, fill=black, draw] (m2) at (2, 0) {};
		\node [rectangle, minimum height=2cm, minimum width=1cm, draw] (D2) at (3.5, 1.25) {$\D_{X, Y_1}$};
		\node [rectangle, minimum height=2cm, minimum width=1cm, draw] (D3) at (3.5, -1.25) {$\D_{X, Y_2}$};
		\node [trapezium, trapezium left angle=0, trapezium right angle=75, shape border rotate=90, trapezium stretches=true, minimum height=1cm, minimum width=2cm, draw] (q) at (6, 1.25) {$q$};
		\node (d1) at (0, -1.75) {}; \node (d2) at (0, 1.75) {}; \node (d3) at (0, .75) {}; \node (d4) at (0, -.75) {};
		\draw [->-] (D1.east |- d3) to [out=0, in=180] node [above, near start] {$X$} (m);
		\draw [->-] (m) to [out=45, in=180] (3.5, 2.75) to [out=0, in=180] node [above, very near end] {$X$} (q.west |- d2);
		\draw [->-] (m) to [out=-45, in=180] node [below, near end] {$X$} (m2);
		\draw [->-] (m2) to [out=45, in=180] (D2);
		\draw [->-] (m2) to [out=-45, in=180] (D3);
		\draw [->-] (D2.east |- d2) to [out=0, in=180] node [below] {$Y_1$} (q);
		\draw [->-] (D3.east |- d4) to [out=0, in=180] node [right, near start] {$Y_2$} (q.west |- d3);
		\draw [->-] (q.east |- d2) to [out=0, in=90] node [above, very near start] {$\RR$} (8, 0) to [out=-90, in=0] (3.5, -2.75) to [out=180, in=0] node [below, very near end] {$\RR$} (D1.east |- d4);
		\draw [->-] (q.east) to [out=0, in=90] (7.5, .5) to [out=-90, in=0] (6, 0) to [out=180, in=0] node [below, very near end] {$\RR$} (D2.east |- m);
		\draw [->-] (q.east |- m) to [out=0, in=90] (7, -.75) to [out=-90, in=0] node [above, very near end] {$\RR$} (D2.east |- d1);
	\end{tikzpicture} \end{center}
	\caption{Hybrid sequential-simultaneous move game}
	\label{fig:upstream-monopolist-open-game}
\end{figure}
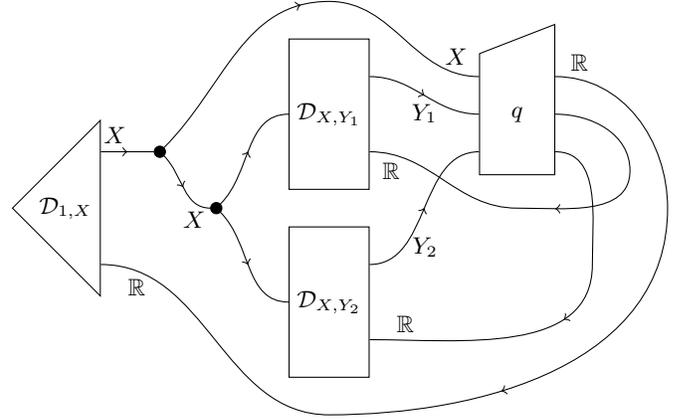

We close this section by combining simultaneous and sequential elements.
Consider the open game in  Figure \ref{fig:upstream-monopolist-open-game}. The game depicts a situation where a player  first makes a
decision ($\D_{1, X}$). This decision is observed by two players who move
simultaneously ($\D_{X, Y_1}$ and $\mathcal D_{X, Y_2}$). As they move simultaneously they cannot observe each others'
moves; they do observe the first player's move though. A possible economic story
 is: two companies use the same input produced by a monopolist, the first player, who sets a
 price for the input. For example the latter players could be rival car manufacturers, and the first player a monopolist who produces tyres. Upon observing the price, both competitors
 decide how much to produce. Profits result accordingly.

The example illustrates how compositionality can be applied to economic
strategic situations. Modelling economic interactions compositionally is very
natural because the object itself is a composition of elements: A market is
composed of competitors, buyers, upstream input providers etc. 
A situation such as this hybrid simultaneous/sequential game is typically modelled as an extensive form game of imperfect information.
We suggest that the representation as an open game, as well as being modular, is a more appropriate representation of the underlying economic situation.

\section{Conclusions and future work}\label{sec:conclusion}

In this paper we have introduced a compositional theory of games that unlike classical
game theory builds larger games from smaller elements. Hence, a game in our theory is
not modelled as one monolithic entity but is constructed from elements which are
glued together by parallel and sequential composition. 
This makes game theory modular.

The categorical tools which allow for compositional games reveal
deep connections to the theory of lenses as well as to the geometry of
interaction. At least to us, these connections were not obvious from the outset
and only later appeared after helpful comments by Jeremy Gibbons and Tom Hirschowitz.

The broader goal of the compositional theory of games started in this paper
is to bring the full force of compositionality to economic modelling. 
In this paper, we have made an important initial step by showing that a compositional theory is possible. 
We have focused only on simultaneous and sequential games, on Nash equilibria as the
solution concept, and have considered only pure strategies. 
Of course, there are many more interesting questions to pursue.

First, an important class of games are repeated games, where players engage in
an interaction more than once, and especially infinitely repeated games. It is well known that these games behave
differently than their one-shot versions. A prime example is the prisoner's
dilemma where in the infinitely repeated version cooperation becomes a possibility for
rational and selfish agents. These games require two innovations: 
a refined equilibrium concept (that is subgame perfect equilibria) as well as more general composition operators. 
See~\cite{ghani_kupke_lambert_forsberg_compositional_treatment_iterated_open_games} for initial results including
the construction of final coalgebras of certain functors on open games which can be used to model infinitely repeated games and thereby bring the powerful concept of bisimulation to bear on infinite games.

Secondly, the solution concept of Nash is built into the definition of an open
game. It is important to consider alternatives. One categorically attractive point of view 
is pursued in \cite{hedges_morphisms_open_games} where open games are taken as objects of a
category and morphisms between them are studied. This is particularly important if we want to define open games, or operators on open games, by universal properties.

Thirdly, applications of game theory very commonly use mixed (probabilistic) strategies, since there are games, such as matching pennies, without a Nash equilibrium in pure strategies. 
Thus, it is important to consider open games with mixed strategies. 
This is possible but surprisingly difficult, and requires some heavier category-theoretic tools, and is work in progress.
A related extension concerns games of incomplete information where some players do not
have access to all relevant informations, for instance other players' utilities.
Coalgebras are a natural way to study this extension \cite{MOSS2004279}.

Lastly, an important practical question is how open games can be implemented and solved.
Calculating with anything larger than a trivial open game by hand is cumbersome and rather error-prone
and the definitions themselves have been developed alongside a prototype Haskell implementation.
Since open games were intended from the beginning to be supported by software tools, 
we have started working with a Haskell based prototype implementation of a string diagram editor, 
a compiler that translates string diagrams into their algebraic expressions, and an engine
that processes the games for simulation, equilibrium checking and other analysis. 

Accordingly, one important application of open games is the compositional development of algorithms to calculate equilibria. 
Various hardness results in algorithmic game theory \cite{Nisan:2007:AGT:1296179} state that approximating solutions of arbitrary games is computationally hard
but this is usually given for classical and `monolithic' games such as normal-form games, without exploiting a formally defined composed structure.
On the other hand, solution algorithms for economic models usually combine various numerical and statistical methods 
like function approximation, root finding, integration and Monte Carlo methods, see for example \cite{winschel:ECTA1029}.
One area where compositional game theory might have a significant impact is in systematically (functorially) combining numerical methods in order to exploit the compositional structure of the game to solve it.

\bibliographystyle{plainurl}
\bibliography{../references}

\end{document}